\documentclass[english,final,twocolumn]{revtex4}
\usepackage[T1]{fontenc}
\usepackage[latin9]{inputenc}
\usepackage{amsmath}
\usepackage{amssymb}
\usepackage{graphicx}

\makeatletter
\@ifundefined{textcolor}{}
{%
 \definecolor{BLACK}{gray}{0}
 \definecolor{WHITE}{gray}{1}
 \definecolor{RED}{rgb}{1,0,0}
 \definecolor{GREEN}{rgb}{0,1,0}
 \definecolor{BLUE}{rgb}{0,0,1}
 \definecolor{CYAN}{cmyk}{1,0,0,0}
 \definecolor{MAGENTA}{cmyk}{0,1,0,0}
 \definecolor{YELLOW}{cmyk}{0,0,1,0}
}

\makeatother

\usepackage{babel}
\begin{document}

\title{New Physics Search with Precision Experiments: Theory Input }

\author{A. Aleksejevs}

\affiliation{Grenfell Campus of Memorial University, Corner Brook, Canada}

\author{S. Barkanova}

\affiliation{Acadia University, Wolfville, Canada }

\author{S. Wu}

\affiliation{Grenfell Campus of Memorial University, Corner Brook, Canada}

\author{V. Zykunov}

\affiliation{Belarussian State University of Transport, Gomel, Belarus }

\begin{abstract}
The best way to search for new physics is by using a diverse set of
probes - not just experiments at the energy and the cosmic frontiers,
but also the low-energy measurements relying on high precision and
high luminosity. One example of such ultra-precision experiments is
the MOLLER experiment planned at JLab, which will measure the parity-violating
electron-electron scattering asymmetry and allow a determination of
the weak mixing angle with a factor of five improvement in precision
over its predecessor, E-158. At this precision, any inconsistency
with the Standard Model should signal new physics. The paper will
explore how new physics particles enter at the next-to-leading order
(one-loop) level. For MOLLER we analyze the effects of dark Z'-boson
on the total calculated asymmetry, and show how this new physics interaction
carriers may influence the analysis of the future experimental results.
\end{abstract}
\maketitle

\section{Precision Parity Violating Physics}

The fact of existence of the Dark Matter \cite{Zwicky-Dark-Matter}
is one of the most striking evidences that the Standard Model (SM)
is incomplete. The further investigation into possible extensions
of SM with new physics particles became one of the main goal of both
theoretical and experimental particle physics. Searches for physics
beyond SM can be summarized into three major directions: energy, cosmic
and precision frontiers. The energy frontier is concentrated on the
direct production of the new physics particles, which might be accessible
at high-energy colliders. In case of the cosmic frontier, direct searches
for new physics are coming from underground experiments, ground and
space telescopes. The precision frontier is driven by the indirect
searches, where new physics particles could impact various observables
in SM and hence cause small deviations from original SM predictions.
This can be studied by using very precise measurements with intense
particle beams. In this paper, we address one of the specific processes
used at precision frontier, namely a test of SM using the parity-violating
Møller ($e+e\rightarrow e+e$) scattering. The most recent parity-violating
Møller scattering experiment, E-158 \cite{E158}, measured parity-violating
right-left asymmetry defined as 
\begin{gather}
{\displaystyle A_{PV}=\frac{\sigma_{R}-\sigma_{L}}{\sigma_{R}+\sigma_{L}}},\label{eq:1a}
\end{gather}
and reported the value of $A_{PV}=(-131\pm14\text{\ensuremath{\pm}}10)\cdot10^{-9}$,
which is resulted in the effective weak mixing angle of $\sin^{2}\theta_{W}^{eff}(Q^{2}=0.026\, GeV^{2})=0.2397\pm0.0010\pm0.0008.$
The reported result is found to be consistent with the SM predictions
(in the $\overline{MS}$ scheme): $\sin^{2}\theta_{W}^{\overline{MS}}(Q^{2}=0.026\, GeV^{2})=0.2381\pm0.0006$
\cite{MSbar-Czarnecki and Marciano,PDG2004}. In order to put more
stringent bounds on the parity violating tests of SM, the MOLLER experiment
planned at the Thomas Jefferson National Accelerator Facility (Jefferson
Lab for short, or JLab) \cite{MOLLER}, will measure $A_{PV}(Q^{2}=0.0056\, GeV^{2})$
at the level of the $\delta(A_{PV})=0.75$ ppb, which translates to
the factor of five improvement in precision for the measurement of
the effective mixing angle compared to the E-158 experiment. At this
level of precision, the new physics signal may be experimentally detectable,
so it is essential to study the potential impact of the new-physics
degrees of freedom on the parity-violating cross section asymmetry
in the Møller scattering.

\section{Dark Photon and Z Bosons}

In our analysis we choose the simplest extension of SM by the additional
$U(1)'$ symmetry proposed in \cite{Holdom}. 

\begin{figure*}
\centering{}\includegraphics[scale=0.3]{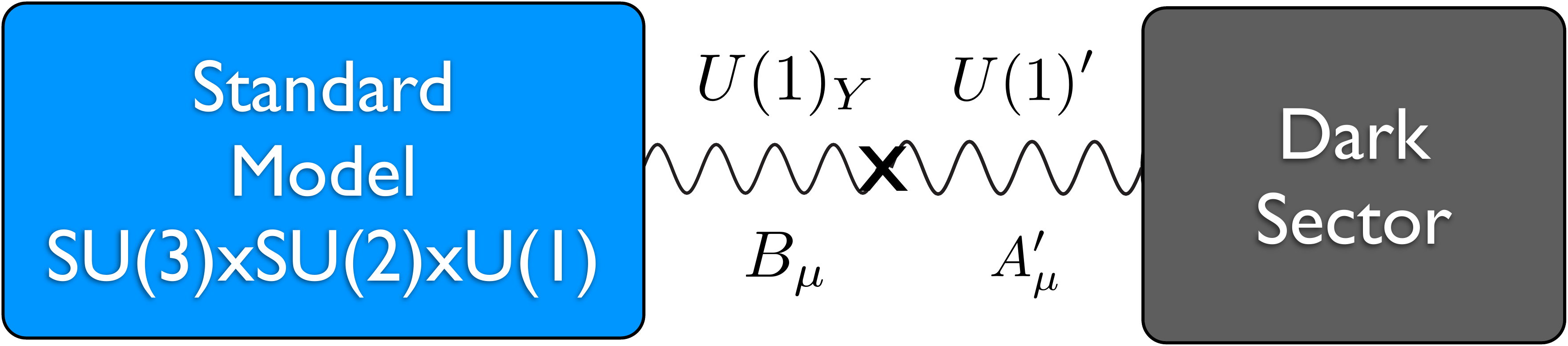}
\caption{Diagrammatic representation of the interaction between Dark Matter
and SM particles through the kinetic mixing between $U(1)_{Y}$ and $U(1)'$.}
\label{fig1}
\end{figure*}

Here, the mixing of $B_{\mu}(U(1)_{Y})$ and $A'_{\mu}(U(1)')$ fields
is expressed through the kinetic mixing Lagrangian (see Fig.\ref{fig1}):
\begin{eqnarray}
\mathfrak{L}_{kin}=-\frac{1}{4}B_{\mu\nu}B^{\mu\nu}+\frac{1}{2}\frac{\epsilon}{\cos\theta_{W}}B_{\mu\nu}A'^{\mu\nu}-\frac{1}{4}A'_{\mu\nu}A'^{\mu\nu},\label{eq:1}
\end{eqnarray}
where $B_{\mu\nu}=\partial_{\mu}B_{\nu}-\partial_{\nu}B_{\mu}$, $B_{\mu}=\cos\theta_{W}A_{\mu}-\sin\theta_{W}Z_{\mu}$
and $\epsilon$ is the $(B_{\mu}-A'_{\mu})$ mixing parameter. If
we assume the simplest scenario for the Higgs sector, which is the
SM Higgs doublet plus the Higgs singlet (used for breaking the $U(1)'$
symmetry and giving mass to $A'_{\mu}$), a Lagrangian describing
interaction between the SM fermions and the dark vector boson $A'_{\mu}$,
photon $V_{\mu}$ and weak $Z_{\mu}$ fields has the following form:
\begin{flalign}
\mathfrak{L}_{int}= & \,-eQ_{f}\bar{f}\gamma_{\mu}f\cdot(V^{\mu}+\epsilon A'^{\mu})-\nonumber \\
\nonumber \\
 & \frac{e}{\sin\theta_{W}\cos\theta_{W}}\bar{f}(c_{V}^{f}\gamma_{\mu}+c_{A}^{f}\gamma_{\mu}\gamma_{5})f\cdot Z^{\mu}.\label{eq:2}
\end{flalign}
Here, $Q_{f}$ is the charge of the fermion in units of $e$. Vector
and axial-vector coupling strengths are defined as follows: 
\begin{flalign}
c_{V}^{f} & =\frac{1}{2}T_{3f}-Q_{f}\sin^{2}\theta_{W}\nonumber \\
c_{A}^{f} & =-\frac{1}{2}T_{3f}\,,\label{eq:3}
\end{flalign}
with $T_{3f}$ defined as fermion's third component of the weak isospin.
The Lagrangian in Eq.\ref{eq:2} has only vector-type coupling of
dark $A'_{\mu}$ to fermions, which is coming from the non-zero kinetic
mixing of $V_{\mu}$ and $A'_{\mu}$ fields. At the leading order,
the kinetic mixing term between $Z_{\mu}$ and $A'_{\mu}$ fields
cancels out with their mass mixing term, so as a result $A'_{\mu}$
does not have the axial-vector type of coupling to fermions in Eq.\ref{eq:2}.
Hence, $A'_{\mu}$ is called a dark photon $V'_{\mu}$ $(A'_{\mu}\equiv V'_{\mu})$,
which resembles a massive photon with the coupling weighted by the
mixing parameter $\epsilon$:
\begin{flalign}
\Gamma_{\mu}^{\bar{f}-V'-f}=-i\epsilon\, eQ_{f}\gamma_{\mu}.\label{eq:4}
\end{flalign}
A possible extension with non-vanishing mixing between dark $A'_{\mu}$
and weak $Z_{\mu}$ was explored in \cite{DLM-Dark-Z} with an additional
mass mixing term described by the mixing parameter $\epsilon_{Z'}=\frac{m_{z'}}{m_{z}}\delta$.
Here, $m_{Z'}$ is the mass of the dark $Z_{\mu}'$ boson and $\delta$
is an arbitrary model-dependent parameter. In this scenario, the interaction
Lagrangian is given by
\begin{flalign}
\mathfrak{L}_{int}= & \,-eQ_{f}\bar{f}\gamma_{\mu}f\cdot(V{}^{\mu}+\epsilon A'_{\mu})-\nonumber \\
\nonumber \\
 & \frac{e}{\sin\theta_{W}\cos\theta_{W}}\bar{f}(c_{V}^{f}\gamma_{\mu}+c_{A}^{f}\gamma_{\mu}\gamma_{5})f\cdot(Z{}^{\mu}+\epsilon_{Z'}A'_{\mu}),\label{eq:5}
\end{flalign}
and, as we can see from above, the dark $A'_{\mu}$ couples to fermions
through both vector and axial-vector interactions, which is similar
to the weak $Z_{\mu}$ coupling. Hence, that type of the dark $A'_{\mu}$
in \cite{DLM-Dark-Z} is called the dark $Z_{\mu}'$ boson ($A'_{\mu}\equiv Z'_{\mu}$).
As a result, the coupling $\bar{f}-Z'_{\mu}-f$ is written in the
following form:
\begin{flalign}
\Gamma_{\mu}^{\bar{f}-Z'-f}=\, & -ie\,\Big(S'_{V}\gamma_{\mu}+S'_{A}\gamma_{\mu}\gamma_{5}\Big),\nonumber \\
\nonumber \\
S'_{V}= & \epsilon\, Q_{f}+\frac{\epsilon_{Z'}c_{V}^{f}}{\sin\theta_{W}\cos\theta_{W}},\nonumber \\
\nonumber \\
S'_{A}= & \frac{\epsilon_{Z'}c_{A}^{f}}{\sin\theta_{W}\cos\theta_{W}}.\label{eq:6}
\end{flalign}
In the case when $\epsilon_{Z'}$ goes to zero, the dark $Z'_{\mu}$
becomes the dark photon $V'_{\mu}$. The coupling in Eq.\ref{eq:6}
is parity-violating by its nature. In our analysis we use left/right
handed (chiral) notation which reflects the nature of the parity-violating
interaction by the simple condition of $g_{L}\ne g_{R}$. Accordingly,
in the chiral basis, 
\begin{flalign}
\Gamma_{\mu}^{\bar{f}-Z'-f}= & -ie(S'_{L}g_{L}\gamma_{\mu}\omega_{-}+S'_{R}g_{R}\gamma_{\mu}\omega_{+}),\label{eq:7}
\end{flalign}
where $\omega_{\pm}=\frac{1\pm\gamma_{5}}{2}$ are chirality projectors,
and $g_{\{R,L\}}=c_{V}^{f}\pm c_{A}^{f}$ are the usual SM right-
and left-handed coupling strengths. The scaling parameters $S'_{\{L,R\}}$
can now be expressed the through mixing parameters as: 
\begin{flalign}
S'_{L}= & \frac{1}{g_{L}}\Big(\epsilon Q_{f}+\delta\frac{m_{Z'}}{m_{Z}}\frac{g_{L}}{\sin\theta_{W}\cos\theta_{W}}\Big)\nonumber \\
\nonumber \\
S'_{R}= & \frac{1}{g_{R}}\Big(\epsilon Q_{f}+\delta\frac{m_{Z'}}{m_{Z}}\frac{g_{R}}{\sin\theta_{W}\cos\theta_{W}}\Big),\label{eq:8}
\end{flalign}
and the condition for the dark $Z'_{\mu}$ becoming the dark photon
$V'_{\mu}$ is given by $S'_{R}g_{R}=S'_{L}g_{L}$, which is satisfied
if either $\delta\rightarrow0$ or $m_{Z'}\ll m_{Z}$. Also, if $S'_{R}=S'_{L}=S'$,
dark $Z'_{\mu}$ boson becomes the ``usual'' SM weak $Z_{\mu}$
boson with modified mass and scaled coupling by $\epsilon_{Z'}=\frac{m_{Z'}}{m_{Z}}\delta$.
The condition $S'_{R}=S'_{L}=S'$ is satisfied if $\epsilon\rightarrow0$. 

\begin{figure}
\begin{centering}
\includegraphics[scale=1.2]{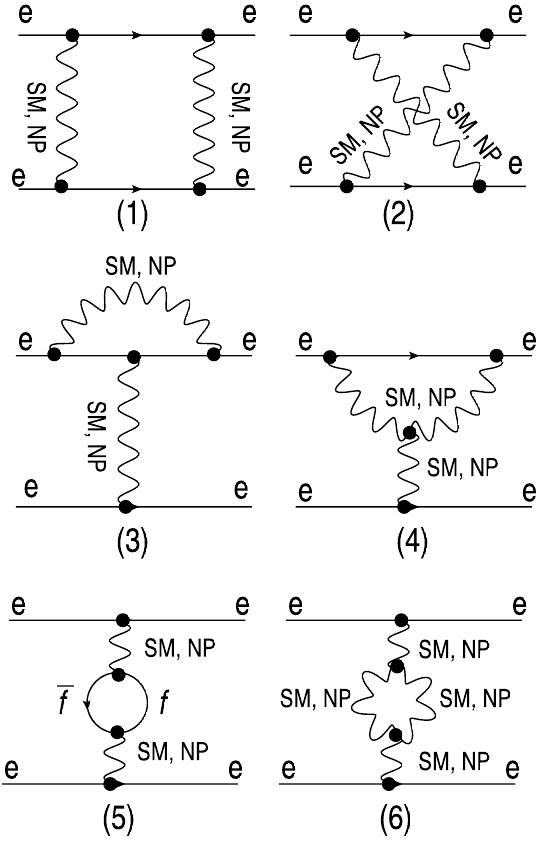}
\par\end{centering}

\caption{Representative one-loop diagrams for the Møller process with the new-physics
(labeled as NP) vector boson in the loops. The label SM stands for
the Standard Model vector bosons. In the actual calculations, the
diagrams with vertex corrections to the lower electron current and
the diagrams for the u-channel are taken into account as well. We
also include the gauge fixing terms in the diagrams with $W^{\pm}$
in the vertex and self-energy graphs (not shown here).}

\label{fig2}
\end{figure}

In this work, we have evaluated the parity-violating asymmetry up
to one-loop level with the dark photon or dark $Z'_{\mu}$ appearing
at the tree level and in the box, vertex, and self-energy diagrams.
Representative diagrams for one loop are shown in Fig.\ref{fig2}.
The diagrams shown in Fig.\ref{fig2} do not contain the Higgs boson
because we do not include the coupling of dark vector $A'_{\mu}$
to the Higgs field, assuming that the diagrams with the Higgs boson
would give a small contribution to the asymmetry. However, for the
sake of completeness, we plan to include this interaction in our next
work. Using on-shell renormalization scheme for SM and NP fields we
have calculated PV asymmetry up to one loop level and included soft-photon
bremsstrahlung when treating infrared divergences. For the SM parameters
we used last-year PDG values. For the cut on energy of the soft-photons,
we choose $\Delta E=0.05\, E_{cms}$ with $E_{cms}=0.106\,\text{GeV}$.

\section{Results and Conclusion}

\begin{figure}
~

~

~

\begin{centering}
\includegraphics[scale=0.31]{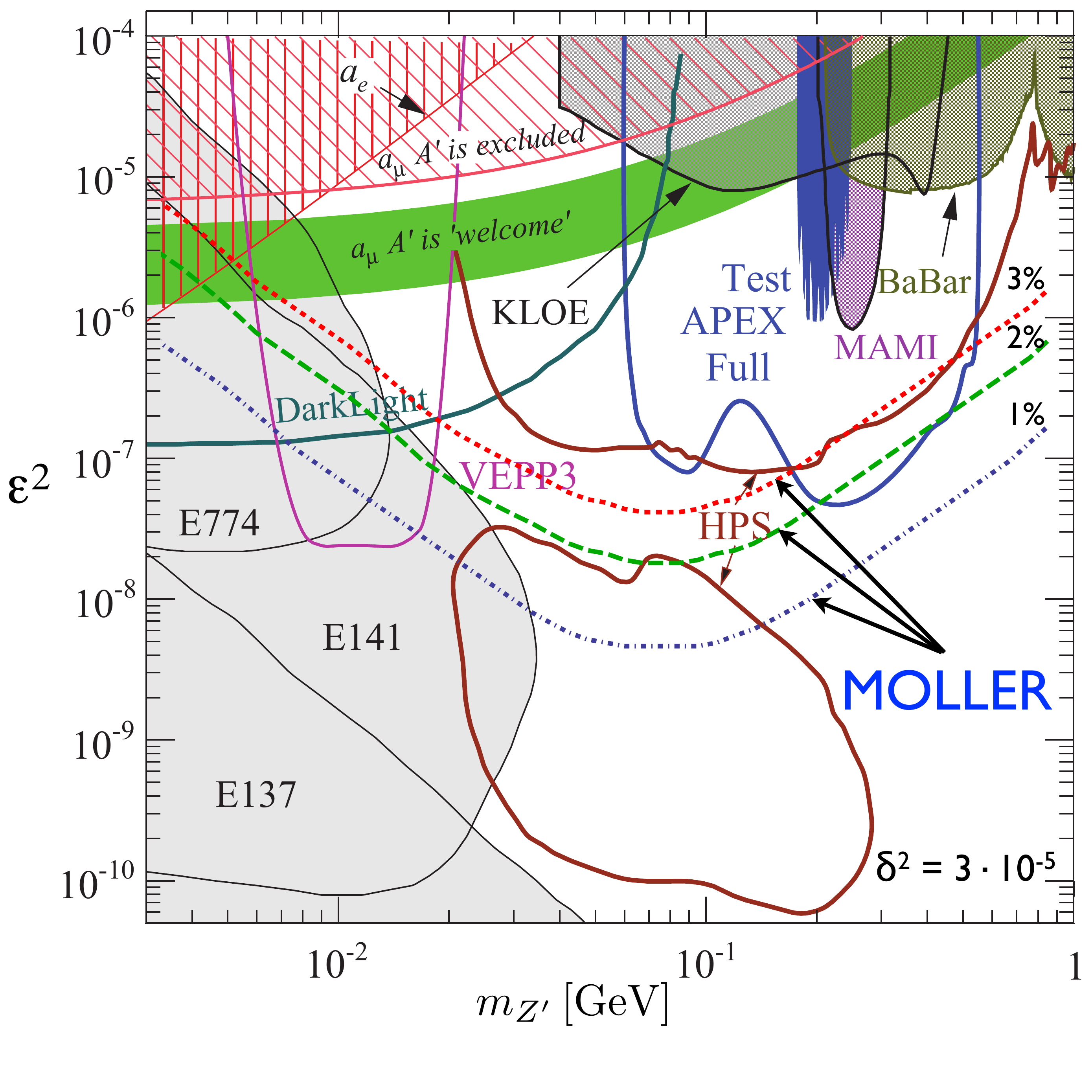}
\par\end{centering}

\caption{Exclusion plots for the dark $Z'_{\mu}$ for the MOLLER experiment
with calculations including one-loop in the on-shell renormalization
scheme, shown against exclusion plot from \cite{New-Physics-Plot-reference}.
We use $\delta^{2}=3\cdot10^{-5}$. The blue dot-dashed, green dashed
and red dotted graphs correspond to 1\%, 2\% and 3\% the PV asymmetry
deviations from the SM prediction, respectively. }

\label{fig3}
\end{figure}

Our calculation strategy basically consist of the following steps.
First, we evaluate the PV asymmetry including one-loop diagrams for
the SM particles. This will determine the SM central value. Then we
proceed with calculations of the PV asymmetry with the new-physics
particles included up to one-loop and construct exclusion plots for
1\%, 2\% and 3\% deviations from the SM central value. Since the MOLLER
experiment is mostly sensitive to the parity-violating interaction,
which is enhanced through the interference term $\sim2\text{Re[}M_{\gamma}M_{Z}]$
in the numerator of Eq.\ref{eq:1a}, we concentrate our attention
on the analysis of dark $Z'_{\mu}$. The exclusion plots for MOLLER
for the case of new physics represented by dark $Z'_{\mu}$ are shown 
in Fig.\ref{fig3}.

In the case if the MOLLER experiment does not detect any significant
deviations from the SM predictions, then this measurement will exclude
everything that is above the corresponding 1\%, 2\% or 3\% lines.
Essentially, if MOLLER does not see the dark $Z'_{\mu}$, it will
exclude the entire region which would explain the $g-2$ anomaly with
the light $Z'_{\mu}$ dark boson. A larger value of the $\delta$
mixing parameter would increase the measurement sensitivity to $Z'_{\mu}$
and push the exclusion lines down. Clearly, as one can see from on
Fig.\ref{fig3}, the MOLLER experiment is very competitive with the
DarkLight \cite{DarkLight}, APEX \cite{APEX}, MAMI \cite{MAMI},
KLOE \cite{KLOE} and HPS \cite{HPS}.

\begin{figure*}
\begin{centering}
\includegraphics[scale=0.5]{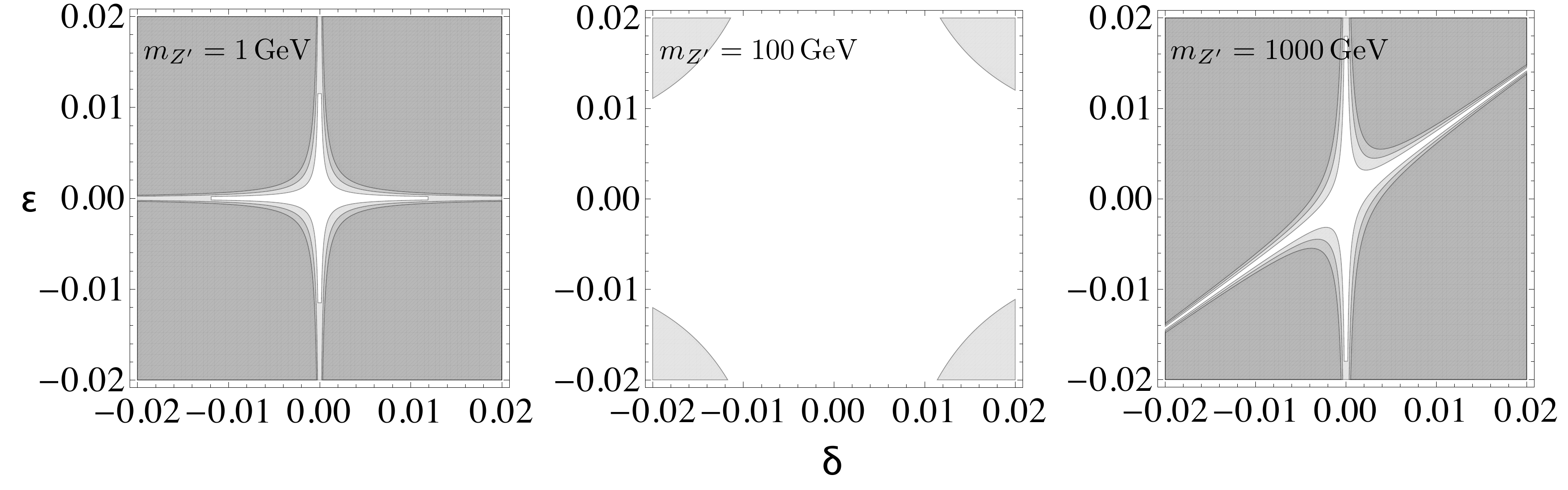}
\par\end{centering}

\caption{Sensitivity of the MOLLER experiment to the mixing parameters $\epsilon$
and $\delta$ for the cases of $m_{Z'}=1,$ $100$ and $1000$ GeV.}

\label{fig4}
\end{figure*}

Fig.\ref{fig4} shows the exclusion regions for the fixed masses of
$Z'_{\mu}$ in the space of $\epsilon$ and $\delta$ mixing parameters.
In the region of the small $Z'_{\mu}$ mass (left plot on Fig.\ref{fig4}),
the overall sensitivity to the variation of $\epsilon$ and $\delta$
is quite high but decreases significantly in the region of the higher
mass of $Z'_{\mu}$ (middle plot of Fig.\ref{fig4}). That is mostly
related to the suppression coming from the dark $Z'_{\mu}$ propagator.
If we assume the scenario of the heavy $Z'_{\mu}$, we observe that
the sensitivity to $\epsilon$ and $\delta$ is enhanced at the leading
order by the term $\sim\frac{\delta}{m_{Z}^{2}}$ and loop contribution
from $Z'_{\mu}$. A detailed analysis of the one-loop contributions
of the dark vector to the PV asymmetry will be addressed in our next
work. In the limit when $\delta\rightarrow0$ (the dark photon), the
sensitivity is weak for all masses of $Z'_{\mu}$. Thus, it is important
to have a non-zero (although possibly small) mixing parameter $\delta$
when it comes to the low-momentum transfer PV experiments such as
MOLLER. In the case of $\epsilon\rightarrow0$ (the ``usual'' $Z_{\mu}$
boson with the modified mass and scaled coupling), we also observe
the reduced sensitivity for the lower masses of $Z'_{\mu}$, so $\epsilon$
should be non-zero in order to satisfy the constrain $|\delta|<1$
(see \cite{DLM-Dark-Z}). For the higher mass of $Z'_{\mu}$ (right
plot of Fig.\ref{fig4}) and the limit when $\epsilon\rightarrow0$,
if no significant discrepancy between the measurement and the SM prediction
is found, we will be able to say that $\delta^{2}\lesssim5\cdot10^{-6}$.
As we can see, for the low-energy frontier, the probability of finding
physics beyond the SM is primarily determined by the level of experimental
precision. Therefore advancing that type of experiments in the precision
domain could actually open a link to our understanding of the nature
of Dark Matter. 

\section{ACKNOWLEDGMENTS}
This work has been supported by the Natural Sciences and Engineering
Research Council of Canada (NSERC). We are grateful to W. Marciano
and J. Erler for the useful discussions and encouragement during the
MITP workshop on ``Low-energy precision physics'' in Mainz in 2013.
Also, many thanks to our undergraduate student research assistants
M. Bluteau, C. Griebler and J. Strickland for testing the first versions
of the code in the summer of 2013. AA and SB  thank JLab Theory Group for 
hospitality during their stay in 2014.

\end{document}